\documentclass[aps,prb,twocolumn,showpacs,floatfix]{revtex4}

\usepackage{graphicx}
\graphicspath{{figs/}}
\begin{document}
\title{Phonon modes at finite temperature in graphene and single-walled carbon nanotubes}
\author{Jin-Wu~Jiang}
    \affiliation{Department of Physics and Centre for Computational Science and Engineering,
             National University of Singapore, Singapore 117542, Republic of Singapore }
\author{Jian-Sheng~Wang}
    \affiliation{Department of Physics and Centre for Computational Science and Engineering,
                 National University of Singapore, Singapore 117542, Republic of Singapore }

%\date{22 December 2009}
\date{\today}
\begin{abstract}
The phonon modes at finite temperature in graphene and single-walled carbon nanotubes (SWCNT) are investigated by the mode projection technique combined with molecular dynamics simulation. It is found that the quadratic phonon spectrum of the flexural mode in graphene is renormalized into linear at finite temperature by the phonon-phonon scattering. Possible influence of this renormalization on the electron-phonon interaction and phonon thermal transport is analyzed. For the SWCNT, however the quadratic property of the phonon spectrum for the flexural mode can survive at finite temperature under protection of the two-fold degeneracy of the flexural mode in SWCNT due to its quasi-one-dimensional structure. The frequency and life time of optical phonon modes are also studied at different temperatures. In particular, the life time of the in-plane optical phonons in graphene and the axial optical phonon in SWCNT can be described by an uniform formula $\tau (T)=560/T$, which coincides with the experimental results.
\end{abstract}

\maketitle

\pagebreak

\section{introduction}
The phonon spectrum in graphene and single-walled carbon nanotubes (SWCNT) is a fundamental and important issue. Phonons can considerably affect electron conduction and thermal transport in carbon-based nano-devices, and the electron-phonon interaction induced Joule heating is one of the most common origins for the breakdown of electronic nano-devices. Considering the importance of phonon modes in graphene and SWCNT, there has been lots of experimental and theoretical works on this topic (for review, see, e.g. Ref.~\onlinecite{review}). In the theoretical models satisfying the rigid rotational symmetry, a special phonon mode with quadratic spectrum can be found which is the so-called flexural mode.\cite{Born,Landau,Mahan,Jiang2006} In graphene, flexural mode is the out-of-plane acoustic phonon mode. It is non-degenerate due to the sheet configuration of the graphene. The flexural mode of SWCNT is a two-fold degenerate transverse acoustic phonon mode. This two-fold degeneracy is due to the quasi-one-dimensional structure of the SWCNT where the two lateral directions are equivalent to each other. The quadratic property of the phonon spectrum ensures the flexural mode to be the lowest-frequency phonon in the long wave length limit, thus it will be the first mode to be thermally excited. As a result of its quadratic phonon spectrum, the flexural mode plays a special role in the electron-phonon interaction. In Ref.~\onlinecite{Mariani}, Mariani {\it et~al.} found that the electron scattering rate from flexural phonon modes will diverge logarithmically with temperature due to the quadratic phonon spectrum of flexural mode. In the ballistic thermal transport in graphene, the transmission function for flexural mode is $T[\omega] \propto \sqrt{\omega}$ due to the quadratic phonon spectrum.\cite{Mingo,Saito,Jiang2009} As a result, the thermal conductance in low temperature region is dominaed by the flexural mode and depends on temperature as $\kappa \propto T^{1.5}$.

Since the qudratic property of the phonon spectrum for flexural mode is important and experiments are carried out at finite temperatures, it is necessary to investigate carefully whether the quadratic property is preserved at finite temperature or not. At finite temperature, phonons can have interaction with electrons, phonons, point defects, and etc. In present paper, we focus on the effect of the phonon-phonon scattering (PPS) on the phonon spectrum. The non-equilibrium Green's function is a standard yet complicate method to study properties of the phonon modes with PPS at finite temperature.\cite{Bonini} In another approach, the PPS can be included in empirical potentials, and its effect on the phonon spectrum can be extracted by molecular dynamics (MD) simulation and velocity correlation function.\cite{Ladd,Maruyama,Wang,McGaughey,Donadio} Using this method in SWCNT, it was found that the flexural mode still has a quadratic phonon spectrum.\cite{Donadio} It will be interesting to investigate the PPS effect on the phonon spectrum in graphene by running MD simulation, since the structures of graphene and SWCNT are essentially different.

In this paper, we investigate the phonon modes at finite temperature in graphene and SWCNT, by using the mode projection technique combined with MD simulation. The carbon-carbon interaction is described by the second-generation Brenner inter-atomic potential.\cite{Brenner} Our results show that the phonon spectrum of flexural mode in graphene is renormalized from quadratic into linear at finite temperature. This renormalization effect should have significant influence in many physical phenomena, since the flexural mode is the first mode to be thermally excited due to its parabolic dispersion. The renormalization does not occur in SWCNT. This difference between graphene and SWCNT is attributed to the two-fold degeneracy of the flexural mode in SWCNT resulting from its quasi-one-dimensional structure. This two-fold degeneracy is robust against the PPS at finite temperature and it can help to protect the quadratic property of the phonon spectrum for flexural mode. We also study optical phonon modes in graphene and SWCNT. The frequency of the out-of-plane optical phonon mode in graphene decreases linearly with the increase of temperature with a slope of $7.5\times 10^{-3}$ cm$^{-1}$/K. The life time of this optical mode is shorter at higher temperature, since the PPS is stronger at higher temperature. For the two degenerate in-plane optical phonon modes in graphene, we find that the frequency shows increasing behavior in low temperature region, and will decrease with further increase of temperature. The frequency reaches a maximum value around 600 K. In wider SWCNT ($n$, $n$) with $n>6$, the axial optical mode behaviors similarly as the in-plane optical mode in graphene. However, in narrow tubes, the frequency of the axial optical mode decreases monotonically with the increase of temperature. The life time of the axial optical phonon mode is not very sensitive to the diameter, and they can be well described by a simple formula for SWCNT with all diameters and graphene as: $\tau (T)=560/T$, which is quite comparable with what the experimentalists have observed.

The rest of the paper is organized as follows. In Sec.~II, we present the detailed calculation procedures. Sec.~III is devoted to the main results  and relevant discussions. The conclusion is in Sec.~IV.

\section{calculation procedure}
\subsection{graphene and SWCNT structures}
\begin{figure}[htpb]
  \begin{center}
    \scalebox{1.2}[1.2]{\includegraphics[width=7cm]{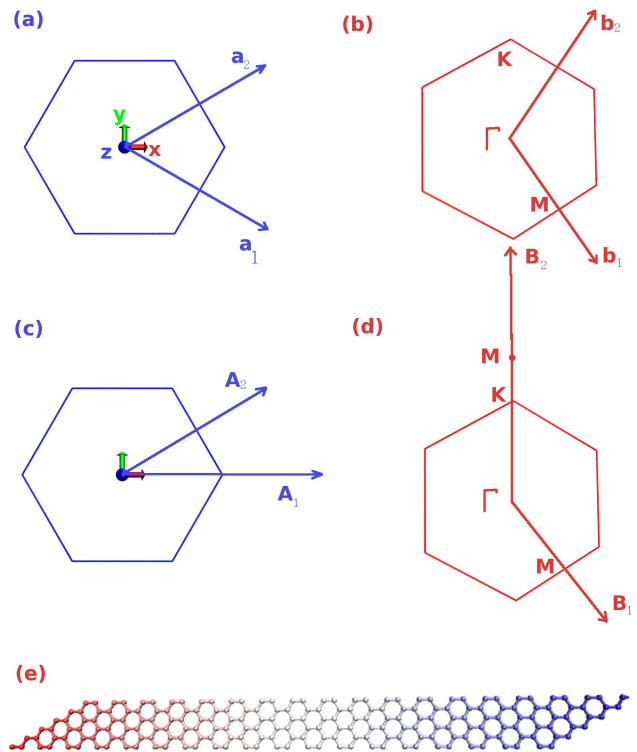}}
  \end{center}
  \caption{(Color online) (a) and (c) are two sets of basis vectors in graphene. (b) and (d) are the corresponding reciprocal basis vectors. A simulation sample of graphene with ($n$, $m$)=(18, 5) is shown in (e).}
  \label{fig_bz}
\end{figure}
The coordinate system in graphene is given in Fig.~\ref{fig_bz}~(a). The two-component basis vectors of the honeycomb lattice structure are\cite{Dresselhaus1998} $\vec{a}_{1}=(\sqrt{3}/2, -1/2)a$ and $\vec{a}_{2}=(\sqrt{3}/2, 1/2)a$, with lattice constant $a=2.46$~{\AA}. Correspondingly, the two basis vectors in the reciprocal space shown in Fig.~\ref{fig_bz}~(b) are $\vec{b}_{1}=(1/2, -\sqrt{3}/2)b$ and $\vec{b}_{2}=(1/2, \sqrt{3}/2)b$, where the reciprocal lattice constant $b=4\pi/\sqrt{3}a$. In lattice structures, there are some arbitrariness in the choice of basis vectors. The only requirement is that the area (two-dimensional lattice) or the volume (three-dimensional lattice) of the unit cell defined by the basis vectors does not change.\cite{White} At the end of this part, we will note that a pair of unsymmetrical basis vectors are more convenient for us, instead of $\vec{a}_{1}$ and $\vec{a}_{2}$ which have reflection symmetry with respect to the $x$ axis. As shown in Fig.~\ref{fig_bz}~(c), we choose basis vectors as $\vec{A}_{1}=\vec{a}_{1}+\vec{a}_{2}$ and $\vec{A}_{2}=\vec{a}_{2}$. The $\vec{A}_{1}$ and $\vec{A}_{2}$ can satisfy the requirement for basis vectors, because $\vec{A}_{1}\times \vec{A}_{2}=\vec{a}_{1}\times \vec{a}_{2}$. Corresponding to $\vec{A}_{1}$ and $\vec{A}_{2}$, the two reciprocal basis vectors are $\vec{B}_{1}=\vec{b}_{1}$ and $\vec{B}_{2}=-\vec{b}_{1}+\vec{b}_{2}$ as shown in Fig.~\ref{fig_bz}~(d).

Fig.~\ref{fig_bz} displays that both lattice and reciprocal lattice of graphene sheet have the same honeycomb structure with $D_{6h}$ group symmetry.\cite{Dresselhaus1998} In the reciprocal lattice space, high symmetry points $\Gamma$, $M$ and $K$ are important in many physical phenomena. Usually, the phonon dispersions along high symmetry lines between these points are shown. The $\Gamma$ point is in the center of the Brillouin zone with zero wave vector. The $M$ point is at $\vec{\Gamma M}=\vec{b}_{1}/2=\vec{B}_{1}/2$. The $K$ point locates at $\vec{\Gamma K}=(-\vec{b}_{1}+\vec{b}_{2})/3=\vec{B}_{2}/3$. Another important wave vector is $\vec{\Gamma KM}=(-\vec{b}_{1}+\vec{b}_{2})/2=\vec{B}_{2}/2$.

A piece of graphene can be denoted by a set of two integers ($n$, $m$). They define two lattice vectors $\vec{R}_{1}=n\vec{A}_{1}$ and $\vec{R}_{2}=m\vec{A}_{2}$, which form two neighboring edges of the graphene. In this graphene sheet, there are $n$ unit cells repeated in $\vec{A}_{1}$ direction and $m$ unit cells in $\vec{A}_{2}$ direction. A total number of $6nm$ phonon modes can be found in this system. By applying periodic boundary conditions, the Bloch theorem guarantees that these phonon modes can be labeled by $n\times m$ wave vectors: $\vec{k}=k_{1}\vec{B}_{1}+k_{2}\vec{B}_{2}$, where $k_{1}=j_{1}/n$, $j_{1}\in(-n/2, n/2]$ and $k_{2}=j_{2}/m$, $j_{2}\in(-m/2, m/2]$. The Bloch theorem connects the wave vector in the reciprocal lattice space to a system structure in the real lattice space. This connection leads to a one to one mapping from the phonon modes we are going to investigate to the shape (also size) of the graphene on which we are going to perform MD simulation.\cite{Born} For example, to study a number of $N_{k}$ phonon modes with wave vectors equally distributed between $\Gamma$ and $M$,
\begin{figure}[htpb]
  \begin{center}
    \scalebox{1.0}[1.0]{\includegraphics[width=7cm]{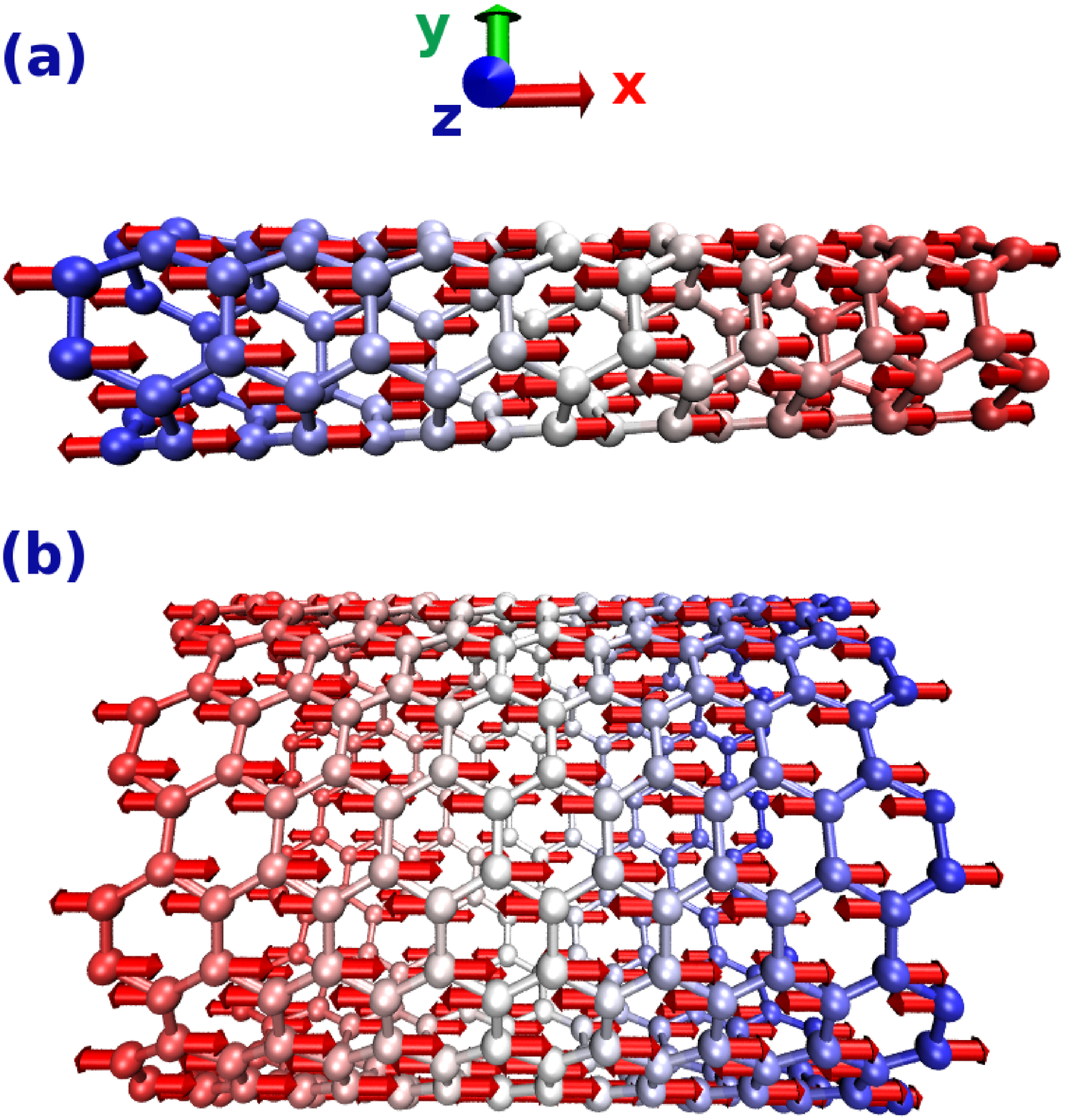}}
  \end{center}
  \caption{(Color online) The vibrational displacement for the axial optical phonon mode in SWCNT (3, 3) and (10, 10).}
  \label{fig_cfg_swcnt}
\end{figure}
 we should investigate a graphene sheet with $n=(N_{k}-1)*2$ and $m=1$. In another situation, suppose that we are going to study a number of $N_{k}$ phonon modes with wave vectors equally distributed along $\Gamma KM$, we have to do simulation on graphene with $n=1$ and $m=(N_{k}-1)*2$. It is apparent that larger system is in need if we want to investigate more phonon modes. Fig.~\ref{fig_bz}~(e) shows the structure of a graphene sheet with ($n$, $m$)=(18, 5). This sample can be used to study those ten phonon modes with wave vectors equally distributed along $\Gamma M$. We have technically set $m$ to be 5 instead of 1, because it should be better to ensure the sizes of the graphene in all directions larger than the interaction range of the Brenner potential.

It is important to note that we can investigate phonon modes along all high symmetry lines (both $\Gamma M$ and $\Gamma KM$) with varying integers ($n$, $m$) by using the two unsymmetrical basis vectors $\vec{A}_{1}$ and $\vec{A}_{2}$. If the two symmetric basis vectors $\vec{a}_{1}$ and $\vec{a}_{2}$ are used, the integers $n$ and $m$ are equivalent to each other; thus only those phonon modes distributed along the high symmetry line $\Gamma M$ can be studied.

The SWCNT ($n$, $n$) is a quasi-one-dimensional system.\cite{Dresselhaus1998} The coordinate system in SWCNT is shown in Fig.~\ref{fig_cfg_swcnt}. We have set the $x$ direction to be the tube axis. SWCNT is periodic in the $x$ direction with basis vector $\vec{A}=a\vec{e}_{x}$, and there are $4n$ carbon atoms in the unit cell. A general lattice vector is denoted by only one integer $l$ as $\vec{R}(l)=l\vec{A}$. Correspondingly, the Brillouin zone is a segment $\vec{B}=2\pi/a\vec{e}_{x}$. An arbitrary wave vector can be described by $\vec{k}=k\vec{B}$, with $k=j/N$, $j\in(-N/2, N/2]$, where $N$ is the number of unit cells. Similar as the situation in graphene, the Bloch theorem requires that if we want to study a number of $N_{k}$ phonon modes with wave vectors $k\in[0, 1/2]$, we need to carry out MD simulation for a SWCNT with the number of unit cell $N=(N_{k}-1)*2$.

\subsection{MD simulation details}
In our simulation, the second-generation Brenner inter-atomic potential\cite{Brenner} is implemented to describe the carbon-carbon interaction. Periodic boundary conditions are applied in all growth directions. The Newton equations of motion are integrated within the fourth order Runge-Kutta algorithm, in which a time step of 0.2 fs is used. The micro-ensemble is applied. Before the MD simulation, the structure is optimized to the equilibrium position at zero Kelvin. Then the initial position and velocity are randomly assigned according to Gauss distribution. $5\times 10^{4}$ simulation steps are used for the system to reach thermal equilibrium state. The trajectory data are recorded for $5\times 10^{5}$ simulation steps. The variation in the temperature is within $2\%$. We have also run MD simulation for $1\times 10^{6}$ simulation steps, and the difference introduced is ignorable. The largest system in our simulation is SWCNT (20, 20) with $N=178$, which has 14240 carbon atoms.

\subsection{mode projection method}
For systems with pure harmonic interaction, phonons are free and have no interaction with each other. As a result, their frequency is temperature independent and the life time is infinite. This is an ideal situation at 0 K, however the PPS is inevitable at finite temperature in real materials like graphene or SWCNT. The PPS leads to the renormalization of the frequency and gives a finite life time for the phonon. Since the PPS is sensitive to the temperature, the phonon frequency and life time will also have temperature dependence. This nonlinear effect on the phonon modes can be investigated by the mode projection method.\cite{Kubo,Wang} Following are the three steps to implement this method, and these are also the three steps in our numerical procedure. We demonstrate this method in graphene. For SWCNT, all processes are the same except some notation substitutions.

(i). We run MD simulation and record the trajectory profile for each atom $(l_{1},l_{2},s)$, say $\vec{r}(l_{1},l_{2},s)$. The two integers $l_{1}$ and $l_{2}$ are lattice indexes for each unit cell in graphene with lattice vector as $\vec{R}(l_{1},l_{2})=l_{1}\vec{A}_{1}+l_{2}\vec{A}_{2}$. The index $s=A, B$ refers to the two nonequivalent carbon atoms in each unit cell. The relative position for $s$ in the unit cell is $\vec{r}_{s}$. The equilibrium position of atom $(l_{1},l_{2},s)$ at 0 K is $\vec{r}_{0}(l_{1},l_{2},s)=\vec{R}(l_{1},l_{2})+\vec{r}_{s}$. As a result, the vibrational displacement can be obtained through $\vec{u}(l_{1},l_{2},s)=\vec{r}(l_{1},l_{2},s)-\vec{r}_{0}(l_{1},l_{2},s)$. With the evolution of MD simulation, time dependent vibrational displacements are achieved $\vec{u}(l_{1},l_{2},s,t)$.

(ii). In periodic lattice structures, the Bloch theorem can be applied\cite{Born},
\begin{eqnarray}
\vec{u}(l_{1}l_{2}s)  =  \frac{1}{\sqrt{N_{1}N_{2}}}\sum_{\vec{k}}\sum_{\sigma}\hat{Q}_{\vec{k}}^{(\sigma)}e^{i\vec{k}\cdot\vec{R}_{l_{1}l_{2}}}\vec{\xi}^{(\sigma)}(\vec{k}|00s),
\end{eqnarray}
where each phonon mode is denoted by two indexes ($\vec{k}$, $\sigma$). $\vec{k}$ is the wave vector and $\sigma$ is the branch index for phonons at same wave vector $\vec{k}$. $N_{1}$ and $N_{2}$ are the number of repeated unit cells in two basis vectors' directions. $\hat{Q}_{\vec{k}}^{(\sigma)}$ is the normal mode operator, and $\vec{\xi}^{(\sigma)}(\vec{k}|00s)$ is the polarization vector for this phonon mode. Considering the orthogonality and completeness of the polarization vectors, the above expression can be reconstructed into another equivalent form:
\begin{eqnarray}
\hat{Q}_{\vec{k}}^{(\sigma)}  =  \frac{1}{\sqrt{N_{1}N_{2}}}\sum_{l_{1}l_{2}s}e^{-i\vec{k}\cdot\vec{R}_{l_{1}l_{2}}}\vec{\xi}^{(\tau)}(\vec{k}|00s)^{\dagger}\cdot\vec{u}(l_{1}l_{2}s),
\end{eqnarray}
where the vibrational displacement $\vec{u}(l_{1}l_{2}s)$ has been obtained in step (i) by MD simulation. In this step (ii), we get the normal mode operator for each phonon ($\vec{k}$, $\sigma$).

(iii). To extract the properties of each phonon mode from its normal mode operator, the normalized auto-correlation function $g(t)$ is calculated:
\begin{eqnarray}
g(t)=\frac{\langle\hat{Q}_{\vec{k}}^{(\sigma)}(t)\hat{Q}_{\vec{k}}^{(\sigma)}(0)^{*}\rangle}{\langle\hat{Q}_{\vec{k}}^{(\sigma)}(0)\hat{Q}_{\vec{k}}^{(\sigma)}(0)^{*}\rangle}.
\end{eqnarray}
For free phonons, this equation simply gives $g(t)=\cos \omega t$, which oscillates without decay indicating an infinite life time. The PPS will lead to finite phonon life time. Under the single mode relaxation time approximation,\cite{Ladd,Maruyama,Wang,McGaughey} we have $g(t)=\cos \omega t e^{-t/\tau}$, where $\omega$ is the renormalized frequency and $\tau$ is the life time.

Briefly speaking, the MD simulation at finite temperature $T$ is carried out to record the information of normal mode operator for each phonon ($\vec{k}$, $\sigma$). The PPS has been considered in MD simulation through the nonlinear Brenner potential. Then the correlation function $g(t)$ is calculated and fitted to $g(t)=\cos \omega t e^{-t/\tau}$. In this way, we obtain the phonon frequency and life time at temperature $T$.

\section{calculation results and discussion}
\subsection{phonon modes in graphene}
\begin{figure}[htpb]
  \begin{center}
    \scalebox{1.0}[1.0]{\includegraphics[width=7cm]{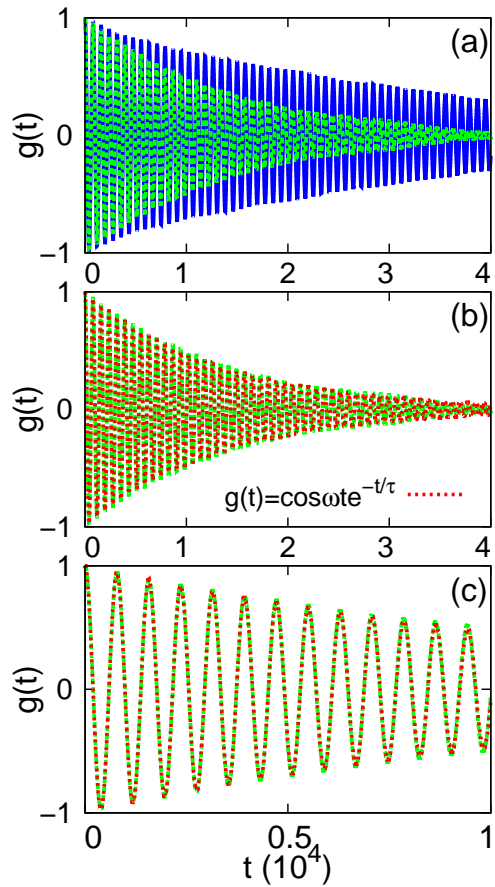}}
  \end{center}
  \caption{(Color online) (a) is the normalized correlation function $g(t)$ at 300 K (blue online) and 1000 K (green online). The $g(t)$ at 1000 K is fitted in (b) and (c).}
  \label{fig_g}
\end{figure}
Fig.~\ref{fig_g} illustrates the process of extracting the phonon frequency and life time from the correlation function $g(t)$. Panel (a) shows correlation functions at 300 K (blue online) and 1000 K (green online). $g(t)$ decays faster at 1000 K than that at 300 K, indicating shorter life time of this phonon mode at higher temperature. Panel (b) is the fitting of the correlation to a function $g(t)=\cos \omega t e^{-t/\tau}$. The fitting procedure is of high accuracy since the fitted curve and the MD simulation results are almost indistinguishable. This means that the single mode relaxation time approximation is quite reasonable. In our numerical fitting procedure, the relative uncertainty for the frequency $\omega$ (life time $\tau$) is about $10^{-6}$ ($10^{-3}$). The first quarter of panel (b) is enlarged in (c) to highlight the details of the fitting.

\begin{figure}[htpb]
  \begin{center}
    \scalebox{1.2}[1.2]{\includegraphics[width=7cm]{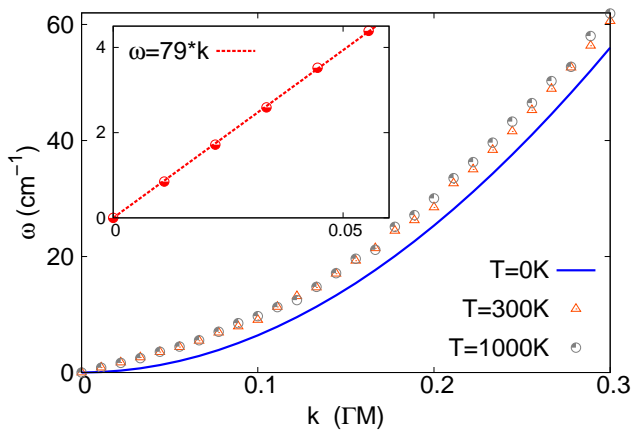}}
  \end{center}
  \caption{(Color online) Renormalization of the flexural mode in graphene. Inset is the linear fitting of the phonon spectrum in long wave length limit.}
  \label{fig_g_fm}
\end{figure}

In graphene, the acoustic phonon spectrum in the $z$ direction shows a quadratic dependence on wave vector $k$: $\omega(k)\propto k^{2}$, which is the so-called flexural mode. Fig.~\ref{fig_g_fm} compares the phonon spectrum of flexural mode at 0, 300 and 1000 K. The phonon spectrum at 0 K is calculated directly by solving the eigen value problem of the dynamical matrix. The phonon spectrum is renormalized obviously, and the curve shows an up-shift at room temperature compared with that at 0 K. Because of this up-shift in the phonon spectrum, it is possible that the dispersion will be renormalized from quadratic into linear. Indeed, such behavior is observed as shown in the inset of Fig.~\ref{fig_g_fm}, which shows the phonon spectrum in the limit of long wave length.
 It shows clearly that the room temperature phonon spectrum can be fitted to a linear function $\omega(k)=ck$ with slope $c=79$, which corresponds to an acoustic velocity of 0.4 kms$^{-1}$ after converting into standard units.

This renormalization of flexural mode at finite temperature can take effect in almost all thermal phenomenon, as it has the lowest frequency and will be firstly excited with increasing temperature. In some physical processes, the quadratic property of the phonon dispersion is important. It is necessary to consider this renormalization effect at finite temperature in those physical processes. For example, the flexural mode has important contribution to the resistance in low temperature region, and the electron scattering rate due to the flexural mode\cite{Mariani} will not diverge with temperature and need to be further renormalized. In the phonon thermal conductance, after the renormalization of the phonon spectrum, the contribution of flexural mode to the thermal conductance will depend on temperature as $\kappa \propto T^{2}$ instead of the $T^{1.5}$.

\begin{figure}[htpb]
  \begin{center}
    \scalebox{1.0}[1.0]{\includegraphics[width=7cm]{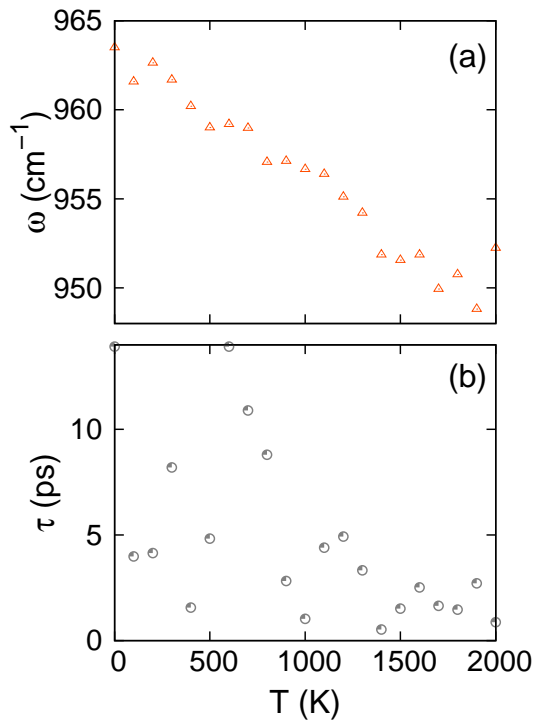}}
  \end{center}
  \caption{(Color online) Frequency and lifetime vs. temperature for the out-of-plane optical phonon mode in graphene.}
  \label{fig_g_op_z}
\end{figure}
\begin{figure}[htpb]
  \begin{center}
    \scalebox{1.0}[1.0]{\includegraphics[width=7cm]{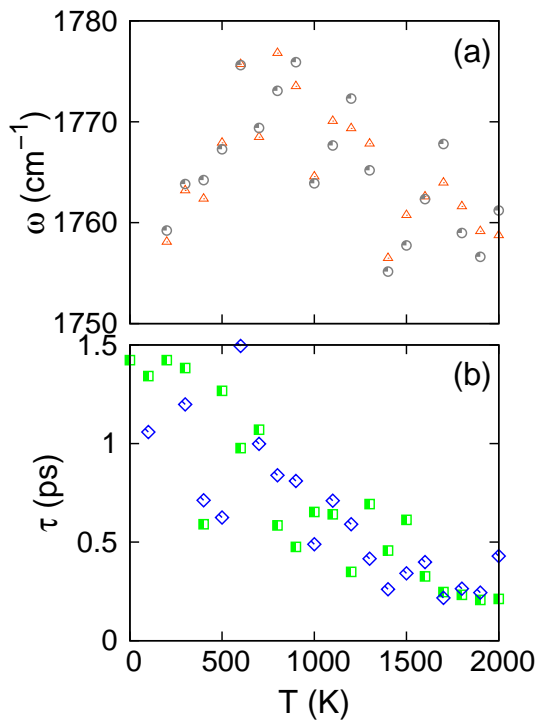}}
  \end{center}
  \caption{(Color online) Frequency and lifetime vs. temperature for the two degenerate in-plane optical phonon modes in graphene.}
  \label{fig_g_op_xy}
\end{figure}
\begin{figure}[htpb]
  \begin{center}
    \scalebox{1.2}[1.2]{\includegraphics[width=7cm]{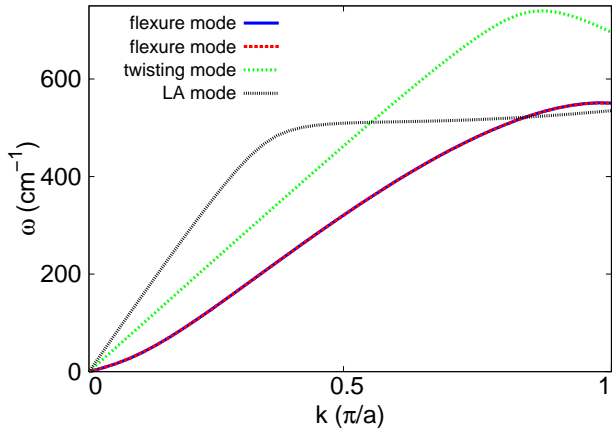}}
  \end{center}
  \caption{(Color online) Phonon dispersion curves for the four acoustic phonon modes in SWCNT (3, 3). The four acoustic phonon modes are: doubly degenerate flexure mode, twisting mode, and longitudinal acoustic (LA) mode.}
  \label{fig_t_acoustic}
\end{figure}

Now we focus on the discussion of the three optical phonon modes at $\Gamma$ point in graphene. Fig.~\ref{fig_g_op_z} shows the frequency and life time of the optical mode in $z$ direction at different temperatures. From panel (a), the frequency decreases monotonically with the increase of temperature and the slope of the decreasing curve is about $7.5\times 10^{-3}$ cm$^{-1}$/K. In Panel (b), an overall decrease in the life time is revealed, however the variation is a little bit for this phonon. Fig.~\ref{fig_g_op_xy} shows the temperature dependence of the frequency and lifetime for the two in-plane optical phonon modes, which are degenerate due to the $D_{6h}$ point group in graphene. The degeneracy of these two optical phonon modes are well preserved, although the value of frequency and life time vary at different temperature. This preservation of degeneracy implies that the $D_{6h}$ symmetry in graphene is protected even at finite temperature. An interesting phenomenon is that the frequency increases with increasing temperature in low temperature region, and it shows decreasing behavior after reaching a maximum value around 600 K. The life time has smaller value at higher temperature, since the PPS is stronger at higher temperature. Our theoretical value (about 1.5 ps at 300 K) is in good agreement with the experiment result, where the life time of this mode is about 1.2 ps at room temperature and is inverse proportional to the temperature.\cite{Kang2010} The agreement between our theoretical results and the experimental value discloses two facts: (i) the PPS dominates the phonon life time under the experimental conditions which is also confirmed by the inverse proportional temperature dependence of life time, and (ii) the mechanical exfoliated graphene samples are of high quality thus other scattering mechanisms are ignorable.

\subsection{phonon modes in SWCNT}
\begin{figure}[htpb]
  \begin{center}
    \scalebox{1.3}[1.3]{\includegraphics[width=7cm]{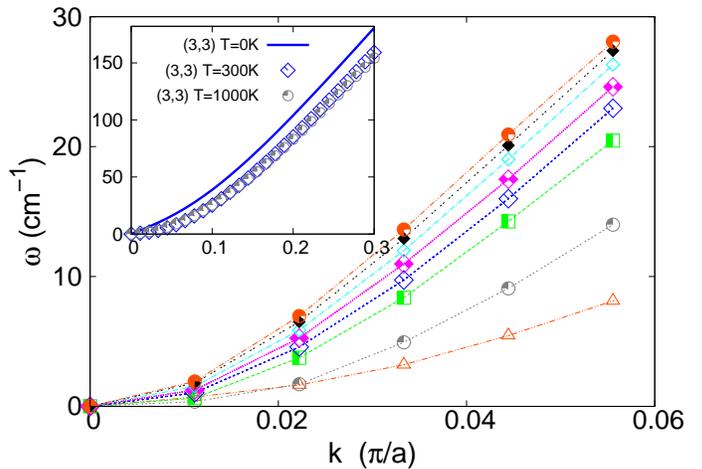}}
  \end{center}
  \caption{(Color online) The flexural mode in SWCNT ($n$, $n$) at room temperature. From bottom to top, $n$=3, 6, 10, 12, 14, 16, 18 and 20. Inset shows the overall figure of the flexural mode in tube (3, 3) at 0, 300, and 1000 K.}
  \label{fig_t_fm}
\end{figure}

In this section, we investigate the phonon modes in SWCNT at finite temperature. The flexural modes in SWCNT are the two transverse acoustic phonon modes, which shows quadratic spectrum at 0 K. They are degenerate because of the quasi-one-dimensional structure where the two vertical directions ($y$ and $z$) are equivalent to each other. Fig.~\ref{fig_t_acoustic} illustrates the four acoustic phonon modes in SWCNT (3, 3): two-fold degenerate flexure mode, twisting mode, and the longitudinal acoustic mode. We focus on the doubly degenerate flexure mode. Fig.~\ref{fig_t_fm} shows the flexural modes in long wave length limit of different SWCNT ($n$, $n$) at room temperature. The phonon dispersion has larger frequency in wider tubes. It is quite interesting that the quadratic property of the phonon spectrum for all tubes studied is not broken by PPS at finite temperature, which confirms previous results.\cite{Donadio} The inset of the figure compares the phonon spectrum for the flexural mode in SWCNT (3, 3) at 0, 300 and 1000 K in a larger range of wave vector.
 It shows that the phonon spectrum has a down-shift at room temperature compared with 0 K. This fact guarantees that the phonon spectrum will not be renormalized into linear dispersion. This result is opposite to what we have observed above in graphene, where the flexural mode is renormalized into linear dispersion at finite temperature. One possible reason for this difference between graphene and SWCNT is that the flexural mode in SWCNT is two-fold degenerate while it is non-degenerate in graphene. The degeneracy in phonon spectrum is robust against PPS at finite temperature, which is also the case for the two degenerate in-plane optical phonon modes in graphene as we have discussed in previous section. This degeneracy may have important effect in the protection of the quadratic property of the phonon spectrum of flexural mode. The degeneracy originates in the quasi-one-dimensional structure of the SWCNT, where a dynamical band gap is opened for the flexural mode due to the four-phonon process.\cite{Martino} This dynamical gap may facilitate the protection of the quadratic phonon spectrum of flexural mode in SWCNT.

\begin{figure}[htpb]
  \begin{center}
    \scalebox{1.0}[1.0]{\includegraphics[width=7cm]{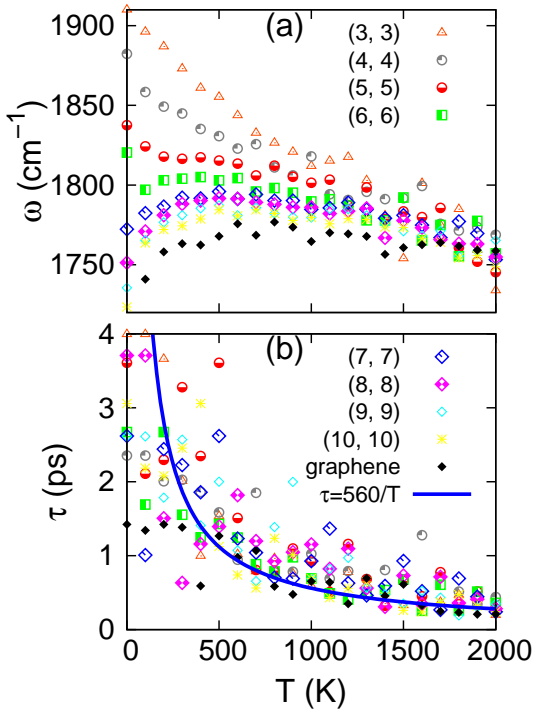}}
  \end{center}
  \caption{(Color online) Frequency and life time v.s temperature for the axial optical phonon mode in SWCNT.}
  \label{fig_t_op}
\end{figure}

Fig.~\ref{fig_cfg_swcnt} exhibits the vibrational morphology of the optical phonon modes in the tube axial direction in SWCNT (3, 3) and (10, 10). In large diameter limit, this mode turns to be the in-plane optical phonon mode in the graphene. The frequency of this mode is shown in Fig.~\ref{fig_t_op}~(a), where the in-plane optical mode in graphene is also shown for comparison. The frequency is smaller in wider SWCNT and approaches to the value of graphene in the limitation of large diameter. In wider SWCNT ($n$, $n$) with $n>6$, the curve behaviors similarly as that shown above for graphene. The frequency increases with the increase of temperature at low temperatures, and then decreases with further increasing temperature above 600 K. The maximum value of frequency is reached around 600 K in all wide tubes. Situation changes in narrow SWCNT where the frequency decreases monotonically with the increase of temperature.
 Fig.~\ref{fig_t_op}~(b) shows that the life time of this mode is not very sensitive to the diameter, and it is roughly inversely proportional to the temperature. All data of life time can be fitted to a function $\tau (T)=560/T$. The experimental value of the life time for this optical phonon at room temperature is around 1.2 ps, varying in samples prepared by different methods.\cite{Song2008,Kang2008} The measured life time is insensitive to the diameter or chiral property of the SWCNT, and it is inverse proportional to the temperature in the experimental range [300, 450]K. The experimental value (1.2 ps) for the life time is smaller than our theoretical results (1.8 ps). This difference is because of other inevitable scattering mechanisms, such as defects or impurities, in the experimental SWCNT samples. Actually, these scattering mechanisms have exposed themselves through the variation in the value of life time for samples prepared by different methods.

\section{conclusion}
To conclude, we study the frequency and life time of different phonon modes at finite temperature in graphene and SWCNT, by using the mode projection technique combined with MD simulation. We pay special attention to the renormalization of the quadratic phonon spectrum of the flexural mode. In graphene, the phonon spectrum of flexural mode is renormalized into linear at finite temperature. The importance of the renormalization is illustrated by some examples. However, the renormalization does not happen in the SWCNT, because of the two-fold degeneracy of the flexural mode in the quasi-one-dimensional SWCNT. For the three optical phonon modes at $\Gamma$ point in graphene, we study the temperature dependence for their frequency and life time. The frequency of the out-of-plane optical phonon mode in graphene decreases with the increase of temperature with a decreasing slope of $7.5\times 10^{-3}$ cm$^{-1}$/K. The two in-plane optical phonon modes in graphene remains degenerate at finite temperature. With the increase of temperature, the frequency will increase to a maximum value around 600 K and then decrease. A similar phenomenon is found for the axial optical mode in wider SWCNT ($n$, $n$) with $n>6$. In narrow tubes, the frequency of the axial optical mode decreases monotonically with the increase of temperature. The life time of the axial optical phonon mode in all SWCNT studied and graphene can be fitted to a function: $\tau (T)=560/T$. This result is in good agreement with the experiments.

\textbf{Acknowledgements} The work is supported by a Faculty Research Grant of R-144-000-257-112 of National University of Singapore.

\end{document}